# Magneto-optical characterization of trions in symmetric InP-based quantum dots for quantum communication applications

**Wojciech Rudno-Rudziński [1,\*], Marek Burakowski [1], Johann Peter Reithmaier [2], Anna Musiał [1], and Mohamed Benyoucef [2,\*]**

[1] Department of Experimental Physics, Faculty of Fundamental Problems of Technology, Wrocław University of Science and Technology, Wybrzeże Wyspiańskiego 27, 50-370 Wrocław, Poland

[2] Institute of Nanostructure Technologies and Analytics (INA), Center for Interdisciplinary Nanostructure Science and Technology (CINSaT), University of Kassel, Heinrich-Plett-Str. 40, 34132 Kassel, Germany

\* Correspondence: wojciech.rudno-rudzinski@pwr.edu.pl; Tel.: +48 71 320 29 86; m.benyoucef@physik.uni-kassel.de; Tel.: +49-561-804-4553

**Abstract:** Magneto-optical parameters of trions in novel large and symmetric InP-based quantum dots, uncommon for molecular beam epitaxy grown nanostructures, with emission in the third telecom window, are measured in Voigt and Faraday configurations of external magnetic field. The diamagnetic coefficients are found to be in the range of 1.5-4 $\mu eV/T^2$, and 8-15 $\mu eV/T^2$, respectively out of plane and in plane of the dots. The determined values of diamagnetic shifts are related to the anisotropy of dot sizes. Trion g-factors are measured to be relatively small, in the range of 0.3-0.7 and 0.5-1.3, in both configurations respectively. Analysis of single carrier g-factors, based on the formalism of spin-correlated orbital currents, leads to the similar values for hole and electron of ~0.25 for Voigt and $g_e \approx$ -5; $g_h \approx$ +6 for Faraday configuration of magnetic field. Values of g-factors close to zero measured in Voigt configuration make the investigated dots promising for electrical tuning of g-factor sign, required for schemes of single spin control in qubit applications.

## 1. Introduction

Self-assembled semiconductor quantum dots (QDs) constitute very interesting objects of studies both from the point of view of fundamental physics and various practical applications [1,2]. Three-dimensional confining potential, resulting in a discreet atom-like levels, predestines them to be a perfect active material in lasers, with some device operation parameters exceeding quantum well counterparts and combined with a broad range of emission wavelengths [3,4]. Moreover, strong confinement of carriers makes them an almost ideal realisation of a two-level system, suitable as a foundation for single photon emitters and other non-classical light sources, also in the spectral region compatible with optical fibre networks [5,6]. On the other hand, due to long spin coherence time [7–9] and possibilities of very fast optical operations, QDs offer a promising playground for spintronic applications, where individual spins of electrons and holes confined in the dots are used instead of charges to store and manipulate information [10,11].

The application of QDs in spintronics requires detailed knowledge on carrier g-factors, determining the susceptibility of the spin to a magnetic field and spin-spin interactions. Various g-factor regimes are practically relevant - for large absolute values the energy of confined states is strongly affected by external magnetics field, leading to considerable separation between states with opposing spins [11]; on the other hand, in the case of g-factor values close to zero, their sign can be changed by e.g. electric field [12–15], which can be used for full Bloch sphere control of the spin qubit [16]. The value of g-factor is strongly material-dependent already in bulk semiconductors [17], leading to large deviations from the value for free electron of $g_e \approx +2$. It is further modified by additional confinement of carriers in nanostructures, which is particularly strong for QDs. Thus, the values of g-factors for a given QD system depend on the details of size, strain and material composition and their distribution within individual dots, as has been recently shown in the picture of spin-correlated orbital currents [18], which results i.a. in the anisotropy of g-factors, e.g. different values of g-factors in the directions perpendicular and parallel to the growth direction [19,20]. It also makes it difficult to predict their values for new nanostructures, requiring experimental verification.

In order to measure g-factors a pump-probe Faraday rotation experiment can be employed, where g-factor value is determined from the Larmor frequency of the spin precession around an external magnetic field. However, this technique has been so far successfully applied only for QD ensembles, giving values averaged over the entire population of dots [21–23]. Addressing individual QD properties requires application of microphotoluminescence spectroscopy (μPL), where g-factors are extracted from the values of Zeeman splitting. As is the case of all the investigations related to QDs, the bulk of work has been devoted to In(Ga)As/GaAs material system [12,14,30,15,21,24–29], with only a very limited number of articles pertaining to InAs dots grown on InP substrates. The latter concern InAsP QDs embedded in an InP nanowire [31] and QDs emitting in the telecom spectral range, either grown by metal-organic vapour phase epitaxy [19,32], or grown by molecular beam epitaxy (MBE), but only with InAlGaAs barriers [20].

In this paper we investigate magneto-optic properties of novel generation of InP-based QDs, with emission at ~1.55 μm, in the centre of the third telecom window. The use of ripening technique during MBE growth of our sample led to nucleation of large and symmetric nanostructures whereas employing InP barriers resulted in considerable intermixing and thus significant phosphorus content in the dots. Such a material system can be used for optically driven spintronics, compatible with telecom infrastructure, but their application potential in this regard has not yet been investigated. In order to fill this gap, we performed the experiments in both Voigt and Faraday configurations, to determine in-plane and out-of-plane characteristics. We focus our studies on charged excitons (trions), because they may be unambiguously identified by magneto-optical spectroscopy and can be used to extract separately electron and hole g-factors in a Voigt configuration. Moreover, they emit light more efficiently than neutral excitons and their emission is always circularly polarised, which facilitates the potential optical manipulation of spin states between different dots. Beside g-factors, we also determine diamagnetic coefficients and relate them to the spatial extents of wave functions, which depend on the structural characteristics of the investigated dots.

## 2. Materials and Methods

### 2.1 Sample growth

QD nanostructures were grown using a Veeco Gen II solid source MBE system on an (100) oriented Fe-doped (*n* type) InP substrate, using two valve solid-source cracker cells for sources of arsenic and phosphorus [33]. QDs were formed by deposition of two monolayers of InAs on InP at a growth temperature of 490°C. The QDs growth was combined with the ripening process [34,35], which resulted in large and high symmetry QDs, with low surface density ranging from 5x10⁸ cm⁻² to ~2x10⁹ cm⁻² [35]. In order to facilitate single dot spectroscopy non-deterministic cylindrical mesas were patterned on the sample surface by means of e-beam lithography and wet chemical etching, where diameters of 1000 nm were chosen for experiments. From preliminary cross-sectional transmission electron microscopy (TEM) scans on the similar buried QDs and atomic force microscopy images made on an analogical sample with surface dots, the dots can be assumed to be lens-shaped, with estimated diameters of 55±15 nm and heights up to 15 nm. Material composition of the QDs is also evaluated from still limited TEM measurements, which show considerable incorporation of phosphorus in dots. For increased extraction efficiency of photons, the QD layer was grown on 25-period InP/InAlGaAs distributed Bragg reflector (DBR), with a refractive index contrast of 0.35. The reflectivity of the DBR reaches around 99% at the telecom C-band [36]. A scheme of the sample layer structure is shown in Figure 1.

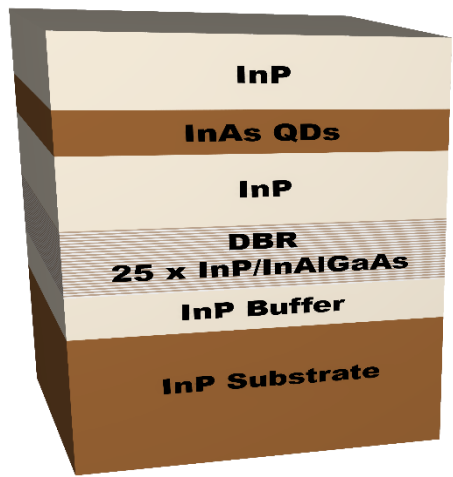

**Figure 1.** Schematic of the investigated sample structure: InP substrate and buffer layer, a 25-pair DBR, and 246 nm thick InP layer below and above nominally two monolayers of InAs forming the self-assembled QDs.

## 2.2 Experimental setup

For spectroscopy measurements the sample was mounted inside the bore of Microstat MO2 from Oxford Instruments, a micro-cryostat with a superconducting magnet providing magnetic field up to 5.0 T, in both Voigt and Faraday configurations. Voigt configuration was achieved with use of an additional mirror inside the bore of the magnet. For all the measurements, the sample was kept at the temperature of 6.0 K. For the microphotoluminescence experiment there was used a continuous-wave semiconductor laser with the emission line at 660 nm, providing non-resonant excitation. Emission from QDs was collected with a microscope objective with NA = 0.4 and 20 mm working distance (20X Mitutoyo Plan Apo NIR Infinity Corrected Objective) and dispersed by a 1 m focal length spectrometer (HORIBA FHR 1000) equipped with an InGaAs linear array detector (HORIBA Symphony II) cooled with liquid nitrogen. This setup provided the spectral resolution of ~20 μeV and spatial resolution of ~1 μm. For the linear polarization-resolved measurements, a rotatable half-wave plate and a polarizer were placed in front of the monochromator.

## 2.3 Nomenclature

The terminology concerning g-factors of quasi-particles in semiconductor nanostructures is not uniform in many publications, which can make them unclear. To make it easier for readers, a brief description of terms used in the results section is provided here. The g-factors determined for the experiments will be identified by subscripts and superscripts, $g_k^l$, with k = e, h or T standing for electron, hole or trion; and l = F or V, standing for Faraday configuration (magnetic field along the growth direction z) or Voigt configuration (magnetic field perpendicular to the growth direction). The sign will be provided where possible, or $|g_k^l|$ notation will be used when only its absolute value can be established. The trion g-factor $g_T^l$ is defined as the algebraic sum of hole and electron g-factors constituting it [37], since it is determined by analysing the evolution in an external magnetic field of energy of optical transitions between an initial trion state, whose influence on the magnetic field depends only on the unpaired carrier g-factor, and a final state being a dot occupied with a carrier left from the pair, of the opposite sign (Figure 2(b)). The values of g-factors for both types of single carriers will be defined by the Zeeman splittings, $\Delta E_{Zeeman} = |g_k^l| \mu_B B_l$, where $\mu_B$=57.9 µeVT⁻¹ is Bohr magneton. Such a definition does not actually agree with a 'classical' relation for the Zeeman splitting of holes $\Delta E_{Zeeman} = 2J_h |g_h^l| \mu_B B_l$ [37], where $J_h$ is the hole spin, since it assumes the hole spin $J_h$ of $\frac{1}{2}$, however, it simplifies and is typically used for the description in the case of QDs, where hole states are composed of heavy-, light- and spin-orbit- states with different spins.

## 3. Results

### 3.1 Trion identification

For a typical MBE grown InAs/InP QD system the degree of anisotropy of dot shape translates into a value of a fine structure splitting (FSS) between two bright states of an exciton in the ground state on the order of at least tens of µeV, which can be clearly observed experimentally. Moreover, since the polarization of emission from the two exciton states is linear and almost orthogonal to each other (as is also in the case of a biexciton), a map of linear polarisation resolved µPL may be used as a tool for distinguishing between charged and neutral excitonic complexes, since in the case of a trion, the ground state is degenerate regardless of symmetry, and emissions from both spin configurations are circularly polarised in an ideal case. However, high symmetry of the QDs investigated here should result in a low value of FSS [36,38], making this method of identification unreliable due to limited spectral resolution of the experimental setup. Figure 2(a) displays linear polarization resolved PL spectra obtained on one of the mesas, exhibiting four intensive and spectrally resolved emission lines that we chose for being representative of a typical behaviour. The results of all the experiments will be illustrated by the examples measured on the same mesa. No energy dependence on the polarization of the emission can be seen for any line in the chosen mesa, as expected for trions or low FSS excitons. However, there can be observed considerable values of the degree of linear polarization, in the range of 15-30%, seen in Figure 2(a) as a change in the PL intensity with the polarization angle. This can be explained by the influence of the valence-band mixing [39], but a detailed analysis of the degree of polarization of the emission is beyond the scope of this work. Since these measurements cannot reveal the origin of emission, other means must be employed. The analysis of the excitation power dependence on the PL intensity, which serves as a standard preliminary scheme of identification of excitonic complexes (results not shown here) yields slightly superlinear dependence, pointing at the direction of trions rather than neutral excitons. However, it is not sufficient to unequivocally distinguish neutral complexes from trions in highly symmetric QDs.

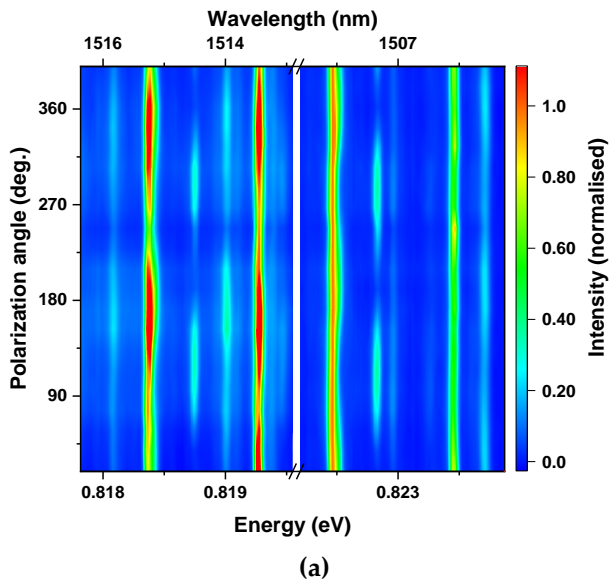

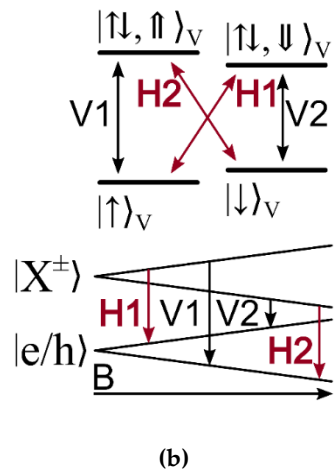

**(a)**

**(b)**

**Figure 2.** (a) Linear polarization-dependent PL spectra in zero magnetic field. (b) Energy level diagram of negative trion states in Voigt configuration. All four optical transitions are allowed, and polarization selection rules are indicated. The evolution of energy vs magnetic field for two groups of transitions with orthogonal linear polarizations (V1 and V2 – 'outer'; H1 and H2 – 'inner') due to Zeeman splittings of initial (trion) and final (dot with single charge) states.

In order to identify the observed excitonic lines, measurements in an external magnetic field in Voigt configuration may be performed. The pattern of energy splitting and polarization of split lines in this case considerably differs between charged and neutral excitons. For trions the emission line splits into an energy quadruplet, which is a consequence of Zeeman splitting of both the initial and final states into doublets, as it is shown in the scheme in Figure 2(b). All four lines have equal intensity due to the field-induced state mixing. All of the transitions are also linearly polarized, with two pairs ('inner' and 'outer') having orthogonal planes of polarization [12,40]. For neutral excitonic complexes, independently from a value of their fine structure splitting, the evolution of energetic structure with magnetic field in-plane is qualitatively different. The dark and bright states of the exciton are always separated by the electron-hole exchange energy and they do not mix with each other [29]. Moreover, the polarization of emission for neutral complexes in high magnetic fields should be circular [29].

Exemplary PL map measured in a magnetic field of 5.0 T in Voigt configuration, plotted as a function of emission energy and an angle of polarization, is shown in Figure 3(a). The lines of interest are separated into triplets composed of two outer lines polarized orthogonally to the internal line. Slight variation from 90° phase difference may be a result of divergence between the strain distribution and the magnetic field induced in-plane anisotropy [39]. In order to make the analysis clearer, Figure 3(b) shows the PL spectra for two orthogonal polarizations. As can be seen, the intensity of the internal line is approximately twice larger than the intensities of each of the outer ones, indicating that it is actually composed of two lines that cannot be resolved in the experiment. This conclusion is also supported by the 30% increase in the broadening of the inner line with magnetic field. Altogether, the obtained results agree with theoretical predictions for the evolution of trion emission, including the observation that the energies of the quadruplet converge for the magnetic field approaching zero (Figure 4). Therefore, it can be concluded that the observed emission indeed originates from trions. The sign of the trion cannot be determined by the optical investigations presented here since for both types of charged exciton complexes the theoretically predicted optical response is qualitatively the same, however *n* doping of the sample suggests negatively charged complexes.

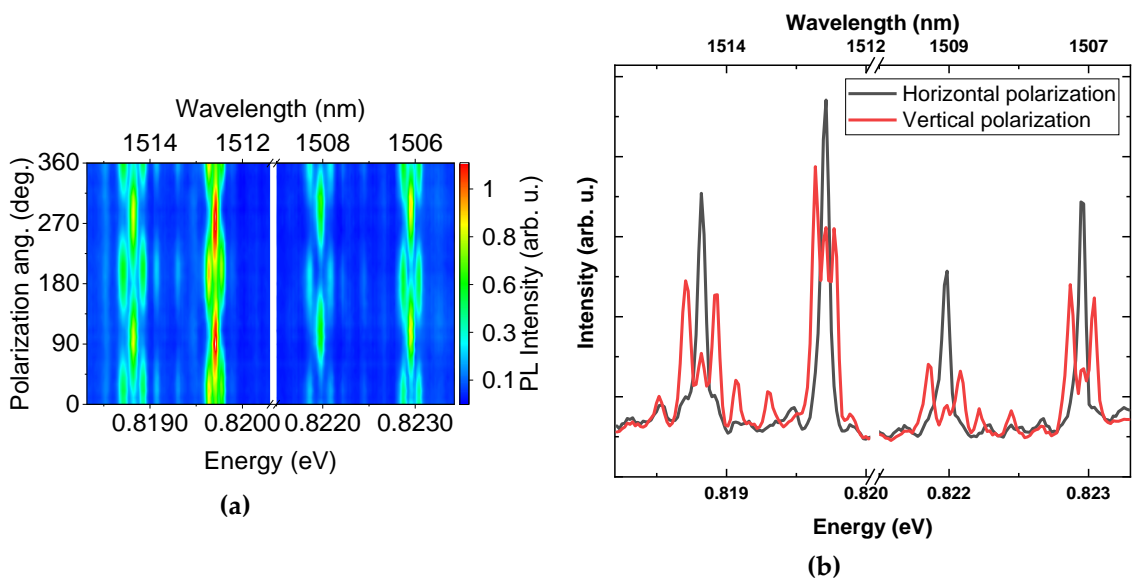

(a)

(b)

**Figure 3.** (a) Polarization-resolved photoluminescence spectra in a magnetic field of B = 5 T in Voigt configuration, showing pattern characteristic for charged exciton emission. (b) PL spectra for two orthogonal polarizations in Voigt configuration at magnetic field of 5.0 T.

### 3.2 Voigt configuration

With a practical tool for trion recognition it was now possible to identify and analyse 15 emission lines related to charged excitons, measured on 5 different mesas. The same four lines, well spectrally isolated and gathered in one spectral window, were chosen to exemplify the results. Figure 4(a) presents evolution of the spectra with magnetic field applied in-plane of the sample, varied in the range from 0.0 T to 5.0 T. Small variations in the relative intensities of emission are attributed to small differences of setup alignment required at each field, indicating that each of the lines originates from a different dot. Since the energy separation between each line of the triplet is larger than the observed linewidths only for the fields above 4.0 T, spectra taken at a range from 4.0 to 5.0 T are shown with smaller step of 0.1 T. Energy dependence of the trion emission on magnetic field in good approximation follows the relations [40]:

$$E_{T,H}(B) = E_T(0) \pm 0.5(g_e^V + g_h^V)\mu_B B + \gamma_T^V B^2,$$

$$E_{T,V}(B) = E_T(0) \pm 0.5(g_e^V - g_h^V)\mu_B B + \gamma_T^V B^2,$$

(1)

where $E_T(0)$ is the trion emission energy without magnetic field, $\gamma_T^V$ is the diamagnetic coefficient and $\mu_B$ stands for Bohr magneton. The second term of the equations 1 is the Zeeman term and describes the spin induced splitting of the trion states. For a detailed analysis and extraction of magneto-optical parameters of investigated QDs the measured PL spectra were fitted with Gaussian functions (due to the influence of spectral diffusion on the spectral lineshape) and obtained energies of the split lines (for the line above 0.822 eV) were plotted as an inset in Figure 4(b). In the next step, the magnetic field evolution of the emission energy of the middle line was fitted with a quadratic dependence, corresponding to the diamagnetic shift, leading to the extraction of a diamagnetic coefficient γ, according to the equation 1, with the value of $\gamma_T^V = 2.0 \frac{\mu eV}{T^2}$ in the presented case. Then, the diamagnetic shift was subtracted from the magnetic field dependence, with the results given in Figure 4(b). A linear dependence on the magnetic field for the two outer lines, corresponding to the Zeeman splitting, can be clearly seen. A fitting procedure using the second term of the equation 1 leads to extraction of the absolute value of the g-factor of the trion $|g_T^V| = 0.74$, taken as an average of g-factors for the two branches of 0.72 and 0.76, since an experiment in the Voigt configuration does not allow sign determination. This value corresponds to the sum of g-factors for a hole and an electron $|g_T^V| = |g_e^V + g_h^V|$. The accuracy of g-factor estimation is on the level of 0.14, basing on the resolution of the setup and fitting uncertainty. On the other hand, since the difference in energy between two internal lines of the quadruplet is below the spectral resolution of the setup, it can be only concluded that the values of $g_e^V$ and $g_h^V$ are very similar, with the difference between them lower than 0.14. Combining both items of information, it can be established that the values of $|g_e^V|$ and $|g_h^V|$ for the investigated dot are very small and equal approximately half of the trion g-factor (~0.37). All the other trion lines were analysed in the same way and showed consistent behaviour.

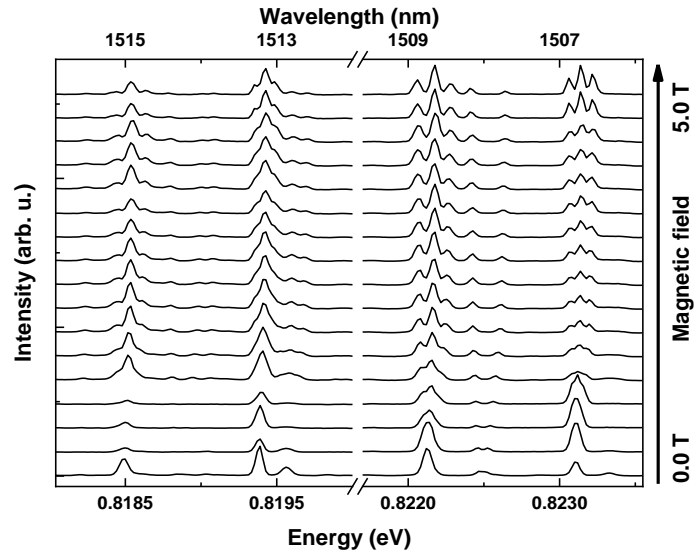

**(a)**

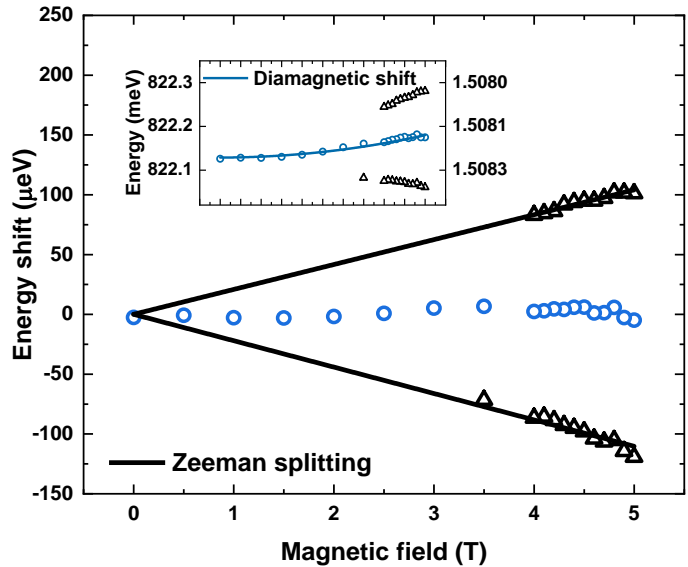

**(b)**

**Figure 4.** Voigt configuration PL spectra. **(a)** PL spectra for magnetic fields from 0 to 4 T with 0.5 T step, and from 4 to 5 T with 0.1 T step **(b)** Energy of optical transitions versus magnetic field, after subtraction of the diamagnetic shift (open black triangles). Black lines are fits to Zeeman energy. Inset shows as measured energy of optical transitions versus magnetic field, without subtraction of diamagnetic shift (blue line).

### 3.3 Faraday configuration

The anisotropy between in-plane and out-of-plane dot structural characteristics (size, shape and composition) should result in differences in values of magneto-optical parameters deduced for orthogonal directions of applied magnetic field [18–20]. In order to obtain such parameters in the growth direction, an experiment in the Faraday configuration was conducted. In this case, magnetic field removes the degeneracy of all Kramer's doublets for trion states [41], resulting in two emission lines separated by the Zeeman splitting and evolving according to the diamagnetic shift:

$$E_T(B) = E_T(0) \pm 0.5|g_T^F|\mu_B B + \gamma_T^F B^2, \qquad (2)$$

where $E_T(0)$ is the trion emission energy without magnetic field, $\gamma_T^F$ is the diamagnetic coefficient and $\mu_B$ stands for Bohr magneton. In the Faraday configuration, the splitting of trion lines depends on the absolute value of the difference between electron and hole g-factors for both types of trions (positive and negative) $|g_T^F| = |g_e^F - g_h^F|$, so the values of g-factors for isolated carriers cannot be determined. Figure 5(a) presents spectra obtained for the exemplary mesa, with increasing magnetic field up to 5.0 T. Diamagnetic shift and splitting into Zeeman doublet are clearly identifiable for all the analysed lines. Analysis conducted to extract the values of diamagnetic shifts and g-factors was performed in a similar fashion as for the Voigt configuration. In the first step, all the transitions were fitted with Gaussian-shaped lines and the obtained transition energies are shown in the inset of Figure 5(b). Then, the average energy of split lines at any given field was calculated and resulting curve fitted with a quadratic function to extract the diamagnetic coefficient. For the line above 0.822 eV, chosen to exemplify the data analysis, the procedure allowed determining the value of $\gamma_T^F = 9.1 \frac{\mu eV}{T^2}$. Finally, the obtained diamagnetic shift was subtracted from the transition energy field dependence and resulting Zeeman split doublet is clearly seen in Figure 5(b). The linear fit lead to the absolute values of trion g-factors in Faraday configuration for the exemplary line of $|g_T^F| = 0.63$.

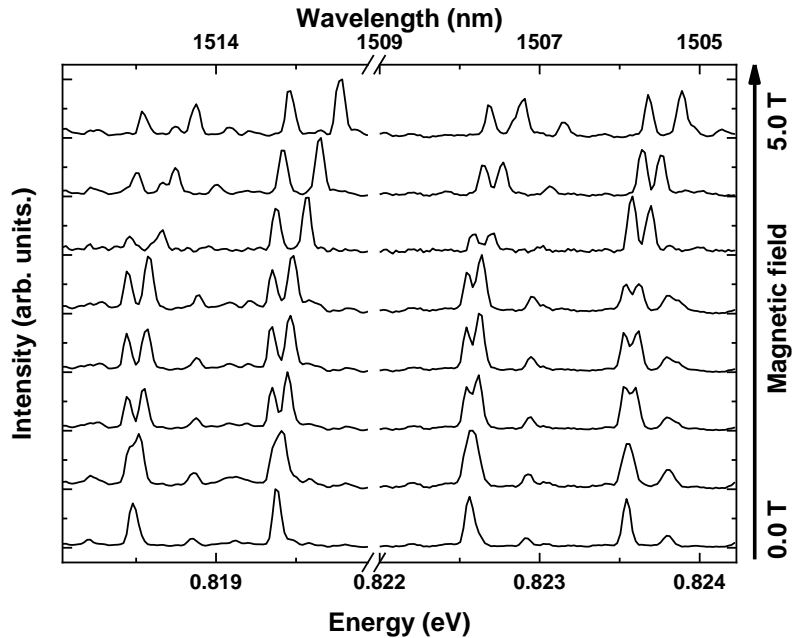

(a)

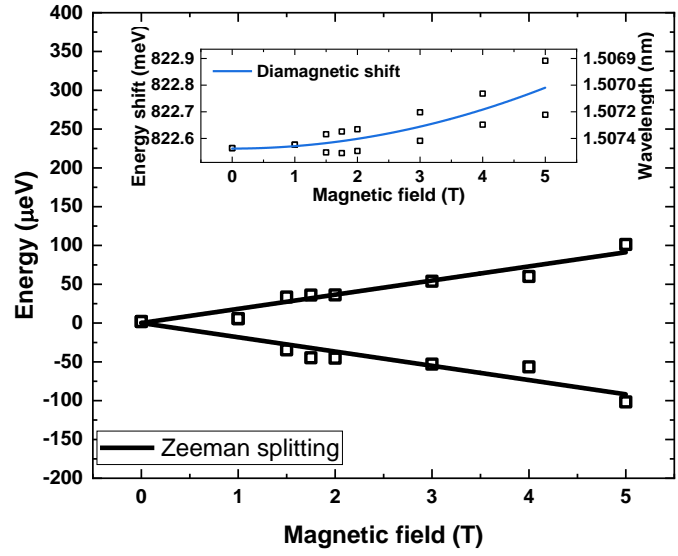

**(b)**

**Figure 5.** Faraday configuration PL spectra. **(a)** PL spectra for magnetic fields from 0 to 5 T. **(b)** Energy of optical transitions versus magnetic field, after subtraction of the diamagnetic shift (open black squares). Black lines are fits to Zeeman energy. Inset shows as measured energy of optical transitions versus magnetic field, without subtraction of diamagnetic shift (blue line).

### 3.4 Summary of magneto-optical parameters

Figure 6 shows the summary of diamagnetic coefficients $\gamma_T^{F,V}$ and trion g-factors $|g_T^{F,V}|$ values determined for all the identified trions, for magnetic field perpendicular and parallel to the growth direction, as a function of trion transition energies. The diamagnetic coefficients are found to be in the range of $\gamma_T^V = 1.5\text{-}4 \ \frac{\mu eV}{T^2}$, centred around $2\text{-}3 \ \frac{\mu eV}{T^2}$, and $\gamma_T^F = 8\text{-}15 \ \frac{\mu eV}{T^2}$, centred around $9\text{-}11 \ \frac{\mu eV}{T^2}$, respectively in Voigt and Faraday configurations. A weak dependence on a transition energy can be seen, where a higher transition energy is attributed to a lower value of the diamagnetic coefficient. Since lower diamagnetic coefficient translates into decreased spatial extent of excitonic complex wave function [42] and thus QD size, and higher transition energy also corresponds to smaller dot sizes, such a dependence can be expected. Further discussion will be presented in Section 4.1. In the case of g-factors for both configurations there is no clear dependence, the spread of sizes, compositions and shapes of individual dots outweighs any systematic relation to transition energies in the relatively narrow range of emission energies available. The determined trion g-factors are found to be in the range of $|g_T^V| = 0.3\text{-}0.7$ and $|g_T^F| = 0.5\text{-}1.3$, respectively in Voigt and Faraday configurations. Detailed analysis and interpretation will be given in Section 4.2.

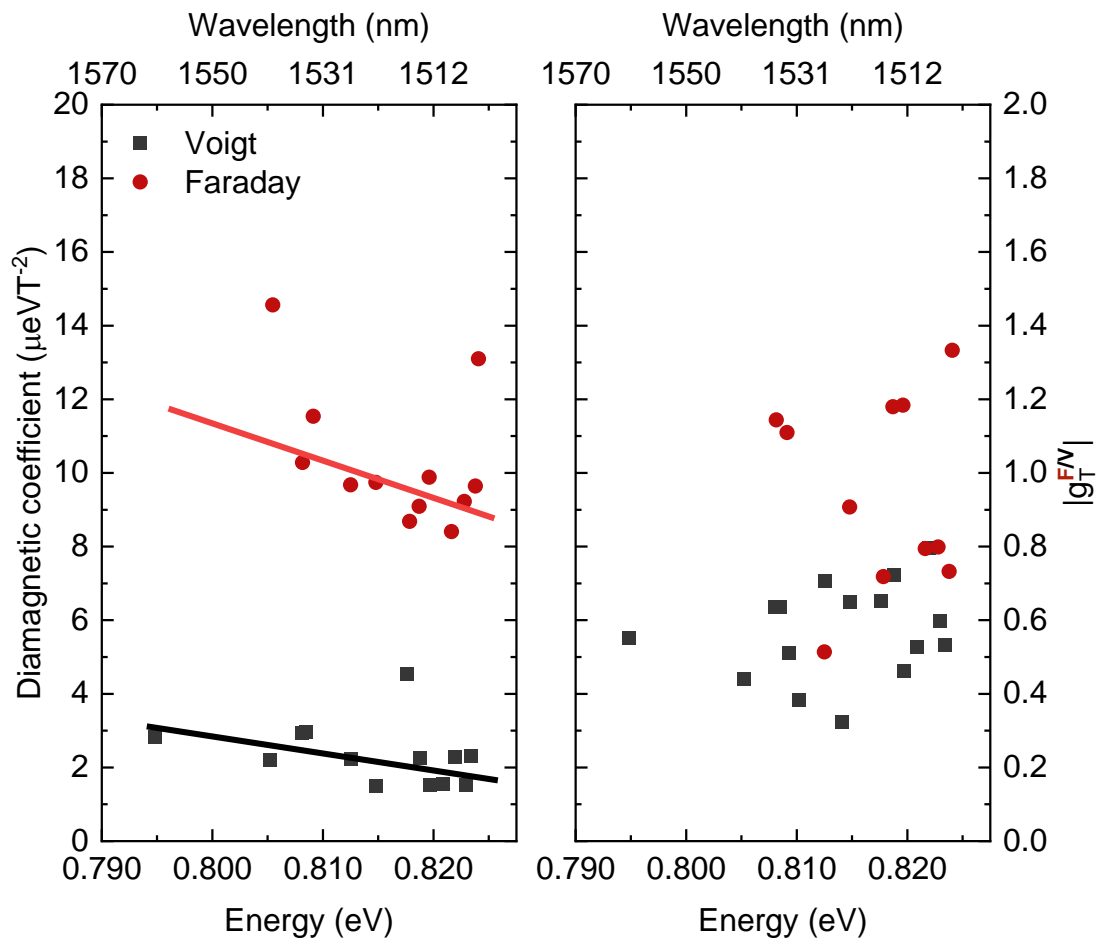

**Figure 6.** Summarized diamagnetic coefficients (left) and g-factors (right) from Voigt (solid black squares) and Faraday (solid red circles) configurations as function of emission energy. Solid lines are guide to the eye.

## 4. Discussion

### 4.1 Spatial extents of wave functions

Generally, the diamagnetic shift of an exciton in a semiconductor structure is determined by the spatial extension of its wave function, however this relation is not straightforward. Nash et al. have shown that in the case of a quantum well a diamagnetic coefficient $\gamma$ is a measure of the in-plane electron-hole separation $\langle \rho_{e-h}^2 \rangle$ and can be expressed as:

$$\gamma = \frac{e^2 \langle \rho_{e-h}^2 \rangle}{8\mu},\tag{3}$$

where $\mu$ is the reduced mass of an exciton [43]. In order to include the influence of additional lateral confinement Walck and Reinecke have employed a generalized gauge transformation which led to expressing the diamagnetic coefficient in two terms related to the size of an exciton – involving confinement and the Coulomb interaction [44]. In the case of QDs with cylindrical symmetry for two extreme instances (either no confinement or no Coulomb interaction), the diamagnetic coefficient is again expressed by equation 3, however the interpretation of the wave function related term changes. In the case when Coulomb interaction is omitted, which is an equivalent of a QD in strong confinement regime, the diamagnetic coefficient reflects only the confinement of an electron and a hole and thus strongly relates to the geometry of a dot and its material composition. There is yet no detailed theory on the relation of diamagnetic shift to the wave function for charged excitons. However, there are publications showing that in quite a broad range of dot parameters the diamagnetic coefficients for trions and excitons in the same dot are practically identical [45]. Theoretical calculations indicate that discrepancies can appear if single particle wave functions of carriers constituting an excitonic complex have considerably different spatial extents [46]. It can happen e.g. for very small dots, where electron wave function leaks to the barrier and wetting layer [47]. However, even then the diamagnetic coefficient for a trion is smaller than for an exciton by only 30% [46].

The dots investigated here are considerably large, therefore the measured coefficients of trions should in good approximation reflect these of neutral excitons. Their optical characteristics indicate that although the ladder of confined states is rather dense, they are still close to the strong confinement regime. Therefore, we transform equation 3 so that it can be used to estimate extents of wave functions in both directions:

$$\langle \rho^2 \rangle = \frac{8\mu\gamma}{e^2}. \tag{4}$$

The exciton reduced effective mass $\mu$ is taken to be 0.039 by using linear interpolation of electron and heavy hole masses in bulk InAs and InP [48]. The calculated values of $\sqrt{\langle \rho^2 \rangle}$ can be understood as standard deviations of squared modulus of a wave function. For Gaussian-like wave functions, which are a good approximation of typical carrier wave functions in QDs, the value of $3\sqrt{\langle \rho^2 \rangle}$ should reflect the total extent of a wave function and be related to the size of confining potential and thus dot shape, size and composition. Although the dot does not have cylindrical symmetry in the direction perpendicular to the growth direction, equation 4 can give some approximation of confinement size also in this case. The application of the abovementioned procedure to the diamagnetic coefficients listed in Section 3.4 gives 5-8 nm in the growth directions and 12-13 nm in plane of the dot, with slight increase with decreasing transmission energy (larger dots), as expected. Although there may be some hole mass anisotropy, the considerable difference between the determined sizes is imposed mostly by the QD geometry, confirming the crucial role of confinement. There may be defined an effective size of the dot, understood as a volume (an ellipsoid) that can by occupied freely by carrier wave functions. For the lens-shaped dots its dimensions should be around 2/3 of the maximal dimensions, i.e. 10 nm high, with 36 nm in-plane. As can be seen the ratio of height to width determined from diamagnetic coefficient anisotropy does not reflect the geometrical ratio. It may be explained by one of the two factors – there may be some additional confinement related to inhomogeneous material composition (migrating phosphorus distribution and indium segregation), which can affect strain and thus piezoelectric field or the dot is large enough in plane that the exciton reaches its 'natural' size, i.e. comparable to the size of exciton in bulk material.

Considering experimental values of diamagnetic coefficients for other QD systems emitting at 1.5 μm, the values obtained for InAs/InGaAlAs quantum dashes, i.e. highly elongated dots with much lower heights (150 nm x 20 nm, 3.5 nm high), are very similar $\gamma_T^V = 4 \frac{\mu eV}{T^2}$, and $\gamma_T^E = 10\text{-}15 \frac{\mu eV}{T^2}$ [49] whereas the values for more symmetric metal-oxide vapour phase epitaxy grown InAs/InP dots with smaller diameter and even lower heights (34 nm diameter, 2 nm high) gave smaller values of $\gamma_T^V = 2 \frac{\mu eV}{T^2}$ [18], and $\gamma_T^E = 7\text{-}10 \frac{\mu eV}{T^2}$ [32] in agreement with difference in the overall geometry of these groups of nanostructures.

*4.2 Analysis of g-factors*

A considerable scattering of the determined trion g-factors, especially in the case of Faraday configuration, is typical for QD studies. It can be related to the sensitivity of hole g-factors to the shape, composition and strain in the individual QD via its strong effect on mixing of the typically predominating heavy hole state with other valence band states, mostly light hole ones [50,51].

In order to analyse the results quantitatively a model developed by van Bree et al. will be used [52]. The magnetic moment of a given carrier has contributions from its spin and orbital degrees of freedom. In the case of carriers confined in semiconductor nanostructures, due to the presence of spin-orbit interaction and coupling between bands, the electron g-factor value may be dominated by the spin-correlated orbital moment $\mu_{ORB}$. For QDs with cylindrical symmetry, a following relation can be derived for a value of the electron g-factor:

$$|g_e^{F,V}| = 2 + 2\frac{\mu_{ORB}^{V,F}}{\mu_B}. \tag{5}$$

Since the orbital moments are generated through orbital currents flowing in the plane perpendicular to the field direction, they are related to the dot geometry in such a way that $\mu_{ORB}^F$ is proportional to the square of the dot radius and $\mu_{ORB}^V$ to the product of the radius and height, thus the ratio of $\frac{\mu_{ORB}^F}{\mu_{ORB}^V}$ is equal to the dot aspect ratio.

In the case of the investigated structures it was found that the absolute values of electron and hole g-factors in Voigt configuration cannot be distinguished experimentally (the inner lines cannot be spectrally resolved) and they are equal to ~0.25. Such proximity to zero facilitates electrical tuning of g-factor signs, which is favourable for some applications. For further analysis it will be assumed that this number is positive. Basing on the determined value of $|g_e^V|$ and equation 5 it is now possible to calculate $\mu_{ORB}^V$ = -0.875 and, taking into account the aspect ratio of investigated dots of around 4, $\mu_{ORB}^F$ = -3.5. The electron g-factor in plane of the dot calculated from equation 5 for such a value of orbital moment is equal to $g_e^F$ = -5. In order to relate this number to the experimentally determined value of the trion g-factor $g_T^F$ the knowledge on the hole g-factor is required. Numerical calculations shown in Reference [42] for the strained InAs/InP dot of 15 nm diameter (smaller than the dots investigated here) and 15 nm height give the value of $g_h^F$ = +6, which combines with $g_e^F$ = -5 for a trion g-factor of 1, in excellent agreement with the value determined experimentally. Although the dot calculated in Ref [42] has smaller diameter and the tendency shown there indicates that for larger dot the hole g-factor could be larger, its increase is limited by the influences of the second-order hole g-factor and light hole mixing, which both decrease the total hole g-factor. In conclusion, the measured values of g-factors seem to be reasonable and relate well to the dots' geometry.

## 5. Summary

Magneto-optical parameters of trions in novel large and symmetric InAsP/InP QDs grown by MBE, with emission in the third telecom window, have been determined in Voigt and Faraday configurations of external magnetic field. The diamagnetic coefficients have been found to be about 2-3 $\frac{\mu eV}{T^2}$ and 9-11 $\frac{\mu eV}{T^2}$ in Voigt and Faraday configurations respectively, slightly increasing with the dot sizes (evidenced by the decreasing emission energy). It is consistent with the interpretation of diamagnetic coefficients as related to the spatial extents of wave functions of emitting excitonic complexes. The determined values of diamagnetic shifts have been related to the dot geometry in plane and out of plane of the dot. The trion g-factors have been measured to be relatively small, in the range of $|g_T^V|$ = 0.3-0.7 and $|g_T^F|$ = 0.5-1.3, respectively in Voigt and Faraday configurations. Analysis of single carrier g-factors, based on the formalism of spin-correlated orbital currents, has led to the values of $g_e^F \approx$ -5; $g_h^F \approx$ +6 for Faraday and $|g_e^V| \approx |g_h^V| \approx 0.25$ for Voigt configuration of magnetic field. Results obtained in Faraday configuration are significantly higher than previously reported experimental values, but are consistent with theoretical predictions for large QDs. Values of g-factors close to zero measured in Voigt configuration make the investigated dots promising for electrical tuning of g-factor sign, required for schemes of single spin control in qubit applications.

**Author Contributions:** Conceptualization, A.M., W.R., and Mo. B; Formal analysis, Ma.B; Investigation, Ma.B and Mo.B; Resources, Mo.B and J.P.R.; Writing—original draft preparation, Ma.B. and W.R.; Software, Ma.B; Supervision, A.M. and Mo. B; Review and editing, A.M. and Mo.B; Funding acquisition, A.M. and Mo.B. All authors have read and agreed to the published version of the manuscript.

**Funding:** This research was funded by the Foundation for Polish Science co-financed by the EU under the ERDF project entitled „Quantum dot-based indistinguishable and entangled photon sources at telecom wavelengths" carried out within the HOMING programme. This work was also financially supported by the BMBF Project (Q.Link.X) and DFG (DeLiCom).

**Acknowledgments:** We acknowledge fruitful discussions on the physics of trions in magnetic field with Michał Gawełczyk, Andrei Kors for his assistance in the MBE growth process. We also acknowledge K. Fuchs and D. Albert for technical support.